\documentclass[aps,pra,twocolumn,amsmath,amssymb,superscriptaddress]{revtex4-2}

\usepackage{comment}

\usepackage{amsmath,amsfonts,amssymb}
\usepackage{physics}
\usepackage{graphicx}
\usepackage{setspace}
\usepackage{tocloft}
\usepackage[dvipsnames]{xcolor}

\newcommand{\CP}[1]{\textcolor{red}{#1}}
\newcommand{\KE}[1]{\textcolor{purple}{#1}}

\cftpagenumbersoff{figure}
\cftpagenumbersoff{table} 

\usepackage{CJK}
\usepackage{mathtools}
\usepackage[normalem]{ulem}
\usepackage{dcolumn}
\usepackage[mathlines]{lineno}
\usepackage{hyperref}
\usepackage{bm}
\usepackage{amssymb}
\usepackage{amsmath}
\usepackage{braket}
\usepackage{color}
\usepackage{xcolor}
\usepackage{soul} 
\usepackage{graphics}
\usepackage{bbm}

\begin{document} 
\title{Seeing through randomness with topological light}

\author{Cade Peters}1
\affiliation{School of Physics, University of the Witwatersrand, Private Bag 3, Wits 2050, South Africa}

\author{Kelsey Everts}
\affiliation{School of Physics, University of the Witwatersrand, Private Bag 3, Wits 2050, South Africa}

\author{Tatjana Kleine}
\affiliation{School of Physics, University of the Witwatersrand, Private Bag 3, Wits 2050, South Africa}

\author{Pedro Ornelas}
\affiliation{School of Physics, University of the Witwatersrand, Private Bag 3, Wits 2050, South Africa}

\author{Andrew Forbes}
\email[email:]{ andrew.forbes@wits.ac.za}
\affiliation{School of Physics, University of the Witwatersrand, Private Bag 3, Wits 2050, South Africa}
\email[Corresponding author: ]{andrew.forbes@wits.ac.za}

\date{\today}

\begin{abstract}
\noindent \textbf{Transmitting structured light robustly through complex random media is crucial in many applications, from sensing to communication. Unfortunately, the spatial structure of light is distorted in such media due to refractive index inhomogeneities that cause multiple scattering, requiring mitigating strategies such as iterative optimisation and adaptive optics. Here, we use topological light to see through random media without the need for any corrective measures. Using skyrmions as our optical topology, we first demonstrate their robustness to randomness using controlled digital random phase masks before showing the universality of the approach with physical samples, from biological tissue to highly scattering materials. We benchmark the invariance of the topology against orbital angular momentum (OAM) and show no modal crosstalk using topology in channels where orbital angular momentum exhibits crosstalk greater than 70\%. With the control in hand, we transmit images encoded into an alphabet of 10 topological numbers and show information transfer with high fidelity in regimes where traditional degrees of freedom, such as OAM, fail. Our work represents an important step towards noise-free transmission through noisy channels with the spatial structure of light without the need for active compensation strategies, opening potential applications in imaging, sensing and communicating with topology.}
\end{abstract}
 
\maketitle


\noindent Improving the transmission of light through complex and random media is a vibrant field of research that has been extensively studied to date \cite{Bertolotti2022imaging, gigan2022roadmap,rotter2017light,mosk2012controlling,cao2022shaping}, crucial for applications such as communication and imaging in real-world environments, e.g., multimode optical fibre \cite{cao2023controlling}, biological systems \cite{yoon2020deep,He2021polarisation,ntziachristos2010going}, optical micromanipulation \cite{cizmar2010situ} and atmospheric channels \cite{peters2025structured}. The challenge is the random distortion to many of light's degrees of freedom, including randomised phases and intensities (scintillation), increased background noise levels and unwanted intermodal coupling. This can lead to a reduced signal-to-noise ratio and an increase in spatial \cite{sheng2007introduction} and temporal \cite{katz2011focusing} modal crosstalk, with deleterious effects on spatial resolution in imaging \cite{yaqoob2008optical,ding2024wavefront} and channel capacity in communication \cite{cox2020structured}.  In sensing, where light's multiple degrees of freedom (DoFs) hold tremendous promise \cite{cheng2025metrology}, probing deeply is severely limited by the corruption of the DoFs themselves.

A number of approaches have been demonstrated to ``see'' (transmit, sense, image, communicate or focus) through randomness, mainly leveraging on pre- or post-correction by inverting the action of the medium itself as captured by its transmission matrix \cite{popoff2010measuring}.  These include iterative optimisation by phase control on dynamic devices such as spatial light modulators to shape the input wavefront for improved focussing \cite{Vellekoop2007} and sensing \cite{thompson2016wavefront,wang2025polychromatic}, correction of the transmitted wavefront for imaging \cite{bertolotti2012non} or the use of guidestar approaches for in-vivo control \cite{horstmeyer2015guidestar,baek2023phase}. Similar approaches work with quantum states \cite{lib2022quantum}, and can benefit from the information capacity of high-dimensional quantum states of light \cite{valencia2020unscrambling,forbes2020scramble}. If the medium is dynamic, threading light through slow changing regions \cite{mididoddi2025threading} or faster wavefront control \cite{nixon2013real} can be used . Time gating, already many decades old and often implemented by nonlinear optics, remains highly effective for scattering media specifically \cite{wang1991ballistic,xu2025high}, while non-scattering complex media can likewise benefit from a modern take on phase conjugation by nonlinear optics \cite{singh2024light,zhou2025automatic}. It is also possible to correct disorder with disorder \cite{horodynski2022anti}, while harnessing learning approaches such as optical neural networks as well as machine learning has improved the speed and efficacy of many approaches \cite{valzania2023online,kupianskyi2024all,liu2023imaging}. Recent approaches have leveraged off the properties of structured light \cite{forbes2021structured}. For instance, light can be structured for arbitrary complex media by creating an eigenmode of the medium, successfully shown for thin scattering \cite{pai2021scattering}, deep tissue imaging \cite{badon2020distortion} and long path atmospheric turbulence \cite{klug2023robust}. 

All the aforementioned approaches are measurement-based, requiring information on the medium, either directly or inferred by its effect on the light, ideally faster than the medium is changing (i.e., real-time). Measurement-free invariance has been shown for specific degrees of freedom to specific channels, such as the non-separability of vectorial light in unitary channels \cite{nape2021revealing}, but this lacks universality. An exciting new direction to explore is the use of topological forms of light \cite{shen2024optical}, with recent progress in simulated conditions, from entanglement decay in quantum systems \cite{Ornelas2024nonlocal,Ornelas2025topologicalrejection} to polarisation degradation with spatially varying retarders in classical systems \cite{wang2024topological} hinting at its potential. Topology, defined as a map from real space to parameter space, is a property of light that will be preserved if the medium can be written as a smooth deformation of the map,. However, which complex random media satisfy this condition is yet unknown \cite{liu2022disorder,chen2025more}, with no experimental evidence under realistic scenarios.  

Here, we demonstrate the resilience of topological light through complex random media, with the concept illustrated in Figure~\ref{fig:Concept}. We first experimentally simulated the action of these channels with a digital optical toolkit and binary phase masks, demonstrating the resilience of a variety of skyrmionic topologies over a large range of distortion strengths. Furthermore, we test the robustness of these states of light through physical scattering samples including biological and inorganic specimens. We observed remarkable resilience of the topology despite the significant distortion experienced by the beams' spatial profile, achieved with no probe or measurement of the samples' structure and optical properties. Finally, by way of an illustrative example, we demonstrate the potential of topological light to encode and preserve information through highly perturbing media and compare its performance to that of other typical forms of structured light for communications. We encode information into a ten letter topological alphabet, demonstrating negligible crosstalk and allowing for almost perfect information retrieval, even in regimes where orbital angular momentum (OAM) exhibits over 80\% crosstalk and very low fidelity information transport.  Our work is the first demonstration of real-world tests of topology, the first to use a topological alphabet for communication, and offers a new route to robust and correctionless transmission of light through complex media.

\section*{Results}
 
\begin{figure}[h]
    \centering
    \includegraphics[width=\linewidth]{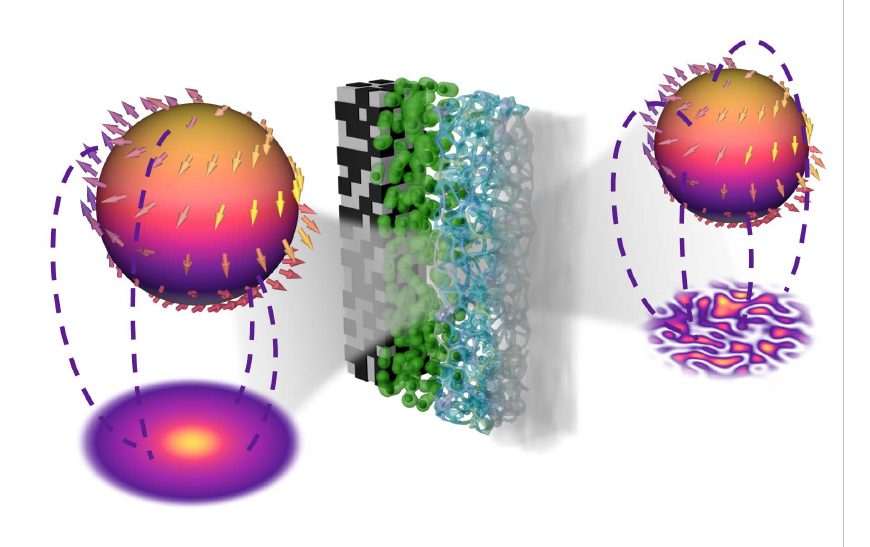}
    \caption{\textbf{Robustness to random media by topology}. Optical skyrmions are topological forms of light defined by a mapping from real-space (the field's traverse spatial profile) to the Poincaré sphere (PS), with the topological texture shown as a stereographic map of vectors on the PS. The topological invariant is the number of times the PS is wrapped.  Many of light's spatial degrees of freedom are distorted when passed through random media, illustrated by the randomised intensity of the light, but the topology persists, leaving the topological wrapping number unchanged.} 
    \label{fig:Concept}
\end{figure}
 
\noindent \textbf{Concept and theory.} Without loss of generality, we consider an optical skyrmion formed in the horizontal $\hat{x}$ and vertical $\hat{y}$ polarisation basis as
\begin{equation}
    U(\mathbf{r}) = LG^0_{l_1}(\mathbf{r})\hat{x} + LG^0_{l_2}(\mathbf{r})\hat{y} \,,
    \label{eq:HV vector beam}
\end{equation}
where $LG_{l}^p(\mathbf{r})$ represents Laguerre-Gaussian (LG) complex field and $\mathbf{r}$ is the transverse spatial coordinate. For convenience we set the radial index $p = 0$ for both fields while allowing the azimuthal index, $l$, to remain non-zero for orbital angular momentum (OAM) of $l\hbar$ per photon. These states of light are commonly referred to as vectorial light and exhibit exotic, spatially varying polarisation structures. Vectorial light is realisable in both the classical and quantum regimes, with mutually relevant principles that are easily transferred from one context to another \cite{shen2022nonseparable,peters2023spatially}. When $|l_1|\neq|l_2|$, these structures form a sphere to sphere mapping, from the transverse plane $\mathcal{R}^2$ to the Poincaré sphere $\mathcal{S}^2$, defining a skyrmionic topology. The topological invariant, $N$, referred to as the skyrmion or wrapping number, uniquely characterises each topology and quantifies how many times one wraps around $\mathcal{S}^2$ after completely traversing $\mathcal{R}^2$ through a stereographic projection \cite{shen2024optical}. The wrapping number for any light field can be calculated using,
\begin{equation}
    N = \frac{1}{4\pi} \int_{\mathcal{R}^2} \epsilon_{ijk} S_i \frac{\partial S_j}{\partial x} \frac{\partial S_k}{\partial y} \text{d}x \text{d}y \,,
    \label{eq:skyrmion wrapping}
\end{equation}
where $\epsilon_{ijk}$ is the Levi-Civita symbol and $S_j$ with $j=1,2,3$ are the locally normalised Stokes parameters such that $\Sigma_{j=1}^3 S^2_j = 1$. This normalisation ensures that the mapping onto $\mathcal{S}^2$ always maps onto a unit sphere. For the above states, Equation \ref{eq:skyrmion wrapping}  simplifies to,
\begin{equation}\label{eq:SkrymeShortcut}
    N = n |l_1-l_2| \,,
\end{equation}
where $n = -1$ if $l_2 > l_1$ and $n=1$ if $l_2 < l_1$. 

When such vectorial fields propagate through complex media, they become aberrated, corrupting many degrees of freedom (DoFs) such as the amplitude, phase, and polarisation structure and OAM spectrum. We may quantify the strength of this distortion with the unitless parameter $\Omega = D/l_0$, where the average distortion imparted onto these fields is solely dependant on $\Omega$ and independent of the type of medium being considered \cite{bachmann2024universal}. Here, $D$ is a characteristic length in the system such as the beam diameter or aperture size and $l_0$ is the correlation length of the medium, quantifying the average distance over which perturbations can still be considered uniform. A smaller $l_0$ corresponds to a larger variation across the medium, resulting in a stronger distortion and thus a larger value for $\Omega$. We may then model an arbitrary, random medium with the use of a binary phase screen $e^{i\Theta(\mathbf{r})}$, where the phase jumps occur over distances equal to $l_0$. If this aberration is applied to a LG beam for example, the distortion can be mathematically described as  \cite{bachmann2024universal,gong2019optical},
\begin{equation}
   LG_l^p (\mathbf{r}) \rightarrow   LG_l^p (\mathbf{r})e^{i\Theta(\mathbf{r})} = \sum_{l,p}c^p_lLG_l^p (\mathbf{r}) \,,
\end{equation}
where $c^p_l$ are complex weighting coefficients. This physically manifests as a heavily distorted amplitude and phase profile. Consequently, the energy (and thus information) that was initially encoded into a single LG mode at the input is randomly and unevenly spread across multiple modes at the output. This results in the loss of information, reducing the performance of communication channels and decreasing the fidelity of imaging techniques.

In contrast, we now show that topology is invariant to the effects of this distortion. To do this, we will examine the effects of $e^{i\Theta(\mathbf{r})}$ on the Stokes parameters $S_j$, which directly influence the topological wrapping number $N$ according to Equation \ref{eq:skyrmion wrapping}. Each Stokes parameter is an observable of one of the Pauli matrices $\sigma_j$ and can be determined by,
\begin{equation}
    S_j = U^{\dagger} \sigma_j U \,,
    \label{eq:Stoke Pauli}
\end{equation}
where $j=0,1,2,3$ and $\sigma_0 = \mathbb{I}_2$. If the input state is perturbed by the phase mask such that $U'(\mathbf{r}) = U(\mathbf{r})e^{i\Theta(\mathbf{r})}$, then the Stokes parameters transform according to,
\begin{eqnarray}
    S'_j &=& U'^{\dagger} \sigma_j U' \nonumber\\
    &=& U^{\dagger}e^{-i\Theta} \sigma_j Ue^{i\Theta} \nonumber \\
    &=&  U^{\dagger} \sigma_j U \,,
\end{eqnarray}
which is the same as Equation \ref{eq:Stoke Pauli}. We see the Stokes parameters remain unchanged under the action of the random medium, even though the underlying fields are significantly distorted. As a consequence, we discover that the skyrmion number must remain invariant according to Equation \ref{eq:skyrmion wrapping}. This is illustrated visually in Figure \ref{fig:Concept}, where a skyrmion beam is heavily distorted in amplitude and phase after passing through a random medium, but the mapping between the transverse plane $\mathcal{R}^2$ and the Poincaré sphere is preserved, leaving the topological wrapping number, $N$, unchanged.\\


\begin{figure*}[htpb!]
    \centering
      \includegraphics[width=\linewidth]{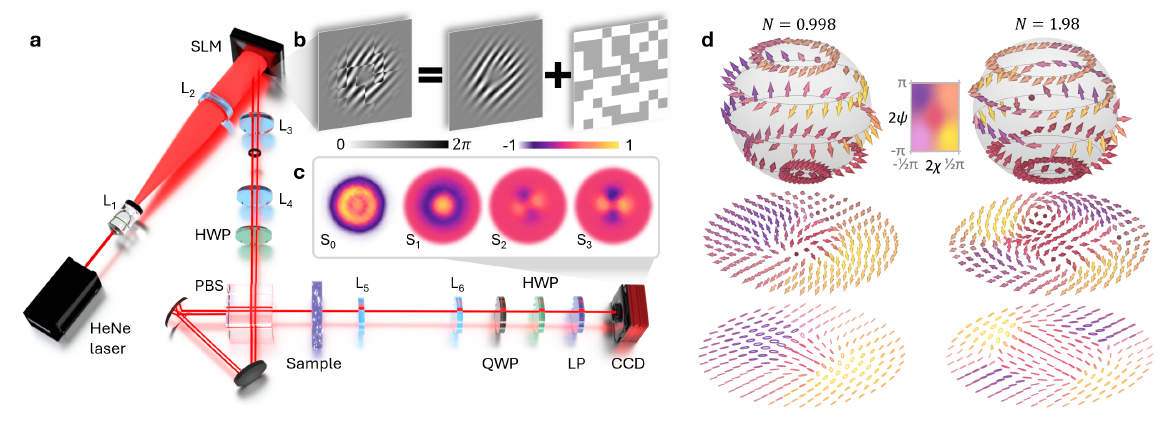}
    \caption{\textbf{Experimentally generating optical skyrmions.} \textbf{a}~Optical skyrmions were generated using an SLM with two holograms, one on each half of the screen, in combination with a modified Sagnac interferometer. The skyrmions were measured using Stokes polarimetry performed with a HWP, QWP, LP and CCD. \textbf{b}~Digital phase holograms were programmed onto the SLM to shape the amplitude and phase of the incident light into LG beams as well as impart the binary phase shifts to digitally simulate the effects of scattering. \textbf{c}~The experimentally measured Stokes parameters of an unaberrated vectorial beam with $l_1=0$ and $l_2=2$. \textbf{d}~The experimental Stokes parameters were used to calculate the Stokes vector field for vector beams with $l_1=1$ and $l_2=0$ (left) and $l_1=2$ and $l_2=0$ (right), giving measured skyrmion numbers of $N = 0.998$ and $N=1.98$, respectively.}
    \label{fig:setup}
\end{figure*}

\noindent \textbf{Experimental creation of optical skyrmions}. In order to verify the robustness of optical Stokes skyrmions through random media, we use the experimental setup depicted in Figure \ref{fig:setup} \textbf{a}. Here, a horizontally polarised, He-Ne laser beam (wavelength $\lambda = 633$~nm) was expanded and then collimated before impinging onto a spatial light modulator (SLM). The SLM screen was divided into two halves and digital phase holograms were programmed onto each half to generate the desired scalar spatial mode using a complex amplitude modulation scheme \cite{arrizon2007pixelated}. An example of a complex amplitude hologram for an $LG^0_3$ beam, a binary phase mask and the combination of the two is shown in Figure \ref{fig:setup} \textbf{b.} The plane of the SLM screen was then imaged with the use of a 4f imaging system, and a spatial filter used to isolate the first diffraction order. At the output of the imaging system, the two beams are converted to diagonal polarisation using a half-wave plate (HWP) and subsequently sent through a Sagnac interferometer to form a vectorial field in the form of Equation \ref{eq:HV vector beam}. A second telescope system was used to image the vector beam onto the detector. The detection apparatus consisted of a HWP, quarter-wave plate (QWP), linear polariser (LP) and charged coupled device (CCD) which allowed for full, spatially resolved, Stokes polarimetry. An example of experimentally obtained Stokes parameters for a vectorial beam of $l_1=0$ and $l_2=2$ is shown in Figure \ref{fig:setup} \textbf{c}. The experimental Stokes parameters were then used to determine the state of polarisation across the beam profile and subsequently the Stokes vector at every point. The experimentally measured polarisation ellipses and the corresponding vector textured fields are shown in Figure \ref{fig:setup} \textbf{d} for states $l_1=1$ and $l_2=0$ (left) and $l_1=2$ and $l_2=0$ (right). We also show their mapping onto the spatial sphere through a stereographic projection, illustrating the skyrmion mapping. The experimentally measured wrapping numbers of $N = 0.998$ and $N = 1.98$ calculated using Equation \ref{eq:skyrmion wrapping} closely match the theoretical values of $N = 1$ and $N=2$, respectively.

This setup was used to investigate the robustness of our skyrmion fields in two ways. First, we performed experiments with the use of digital random phase masks, allowing for precise control over both the incident beam and correlation length of the medium, with further details provided within the Supplementary Information. We studied the structure of vectorial fields with diverse topological structures under the influence of perturbations across a wide range of strengths. Providing an exhaustive study of their topological resilience under random phase perturbations. Second, we tested the robustness of these same fields in a more realistic setting, using a variety of physical inorganic and biological scatterers. These samples allowed for a more accurate evaluation of the robustness of optical skyrmions in real-world applications and complement the digital results allowing us to validate their accuracy at simulating real-world channels.


\begin{figure*}[t!]
    \centering
     \includegraphics[width=0.9\linewidth]{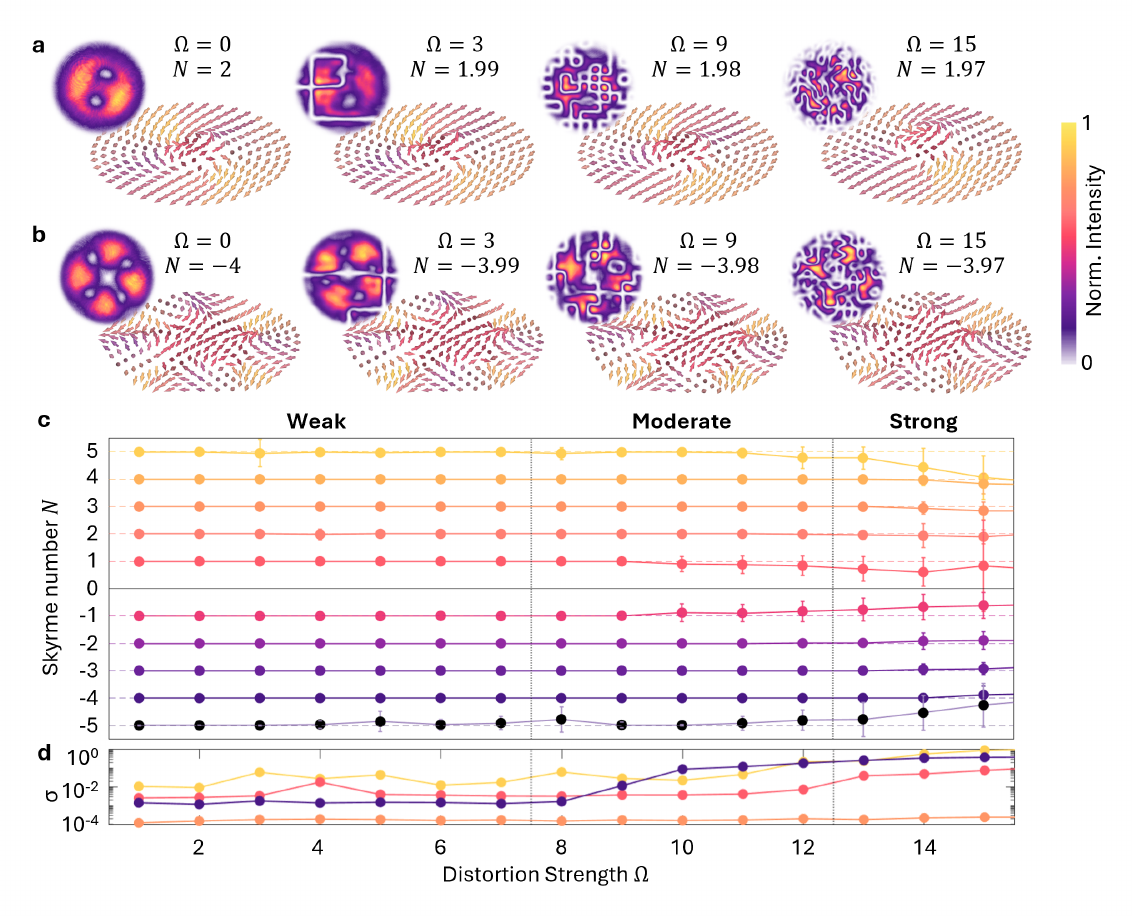}
    \caption{\textbf{Skyrmions through simulated random media.} \textbf{a}~Diagonally polarised intensity projections of the measured skyrmion beams and experimentally reconstructed vector textures for $N=2$ at distortion strengths $\Omega = 0,3,9$ and $15$ for a single realisation of the binary phase mask. \textbf{b}~Diagonally polarised intensity projections of the measured skyrmion beams and experimentally reconstructed vector textures for $N=-4$ at distortion strengths $\Omega = 0,3,9$ and $15$ for a single realisation of the binary phase mask. \textbf{c}~Experimental results showing the average measured skyrmion number $N$ for 10 distinct topologies over 15 distortion strengths $\Omega$. Each data point shows the average over 100 random realisations of the binary phase screens, with the error bars indicating the standard deviation $\sigma$. The distortion strengths are categorised into 3 regimes (weak, medium and strong) based on the extent to which they aberrate an incident plane wave. \textbf{d}~The standard deviation of the measured skyrmion number as a function of distortion strengths for selected input topologies, $N=5,3,1$ and $-4$. }
    \label{fig:resultsdigital}
\end{figure*} 

\noindent \textbf{Digital random phase masks.} Figure \ref{fig:resultsdigital} \textbf{a} and \textbf{b} shows the Stokes vector field and associated diagonally polarised intensity component (shown as an inset) for vector fields with $l_{1,2} = 2,0$ and $l_{1,2} = 0,4$ , respectively, measured under ideal experimental conditions, $\Omega=0$ and 3 different distortion strengths ranging from weak, $\Omega=3$, moderate, $\Omega=9$ and strong, $\Omega=15$. The increasing distortion strength manifests clearly as a distortion of the intensity components of the field (with the diagonally polarised component shown here as an example) as well as the Stokes vector field. The impact of the binary phase on the intensity is clearly evident, a consequence of the high-spatial frequencies as a result of the abrupt phase jumps between $0$ and $\pi$. These manifest as intensity and phase variations after propagation through the finite apertured optical system. Despite this additional distortion of the field (i.e., not only the phase distortion but the amplitude distortions induced by the phase distortion), the topology is recovered successfully in each scenario, even when the field has been distorted beyond recognition. 

Next, we use our digitally implemented random masks on the SLM to study controlled randomness, as well as encode fields with different $l_1$ and $l_2$, for a comprehensive investigation of the skyrmionic topology of our vectorial fields under a full set of random phase perturbation scenarios. The exact combinations of $l_1$ and $l_2$ can be found in the Supplementary Information. A summary of these results are shown in Figure \ref{fig:resultsdigital} \textbf{c} where 10 fields with unique skyrmionic topologies were studied under the influence of phase perturbations with distortion strengths ranging from weak, $\Omega = 1 $, to strong, $\Omega = 15$. To avoid biasing the results towards any particular configuration of random phase mask iterations, every reported skyrmion number measurement is taken as an average over 100 random realisations. The standard deviation over the full set of random mask realizations is shown as error bars for each measured number and explicitly shown on a logarithmic scale as a function of the distortion strength in Figure \ref{fig:resultsdigital} \textbf{d} for $N=5,3,1$ and $-4$. In the weak, regime the average measured skyrmion number perfectly corresponds to the encoded value for all topologies and distortion strengths. Furthermore, the standard deviations remain almost two orders of magnitude below the measured averages, indicating an extremely stable topological structure over the entire range of investigated strengths. In the moderate regime, we see that for all but two of the encoded topologies, the average measured skyrmion number maintains its close agreement with the encoded number. The standard deviation remains consistently below $10^{-1}$ for all topologies, indicating that the variations in the skyrmion number over all of the realizations is unlikely to lead to an incorrect identification to within the closest integer. In the strong regime, we start to observe a degradation of the higher order topological structures namely those with $N=\pm5$. As shown in Figure \ref{fig:resultsdigital} \textbf{a} and \textbf{b} for $\Omega=15$, the intensity distributions of the fields become completely unrecognizable. This results in larger possible discrepancies when measuring the skyrmion number which relies on the identification of polarisation singularities embedded within the polarisation field $S_2 + iS_3$. Higher order topologies are comprised of more polarisation singularities. This means that by increasing the perturbation strength, the likelihood of not measuring a particular singularity or measuring a false singularity increases. A in a change of the measured skyrmion number by a magnitude of $0.5-1$ for each realisation is seen as a result. We expect $\Omega = 15$ to be the limit at which the topological structure starts to become challenging to accurately measure. AT this point, the standard deviation for the higher-order topologies starts to reach $1$, indicating that for individual realisations of noise the topology may be falsely identified.

\begin{figure*}[htpb!]
  \centering
   \includegraphics[width = 0.9\linewidth]{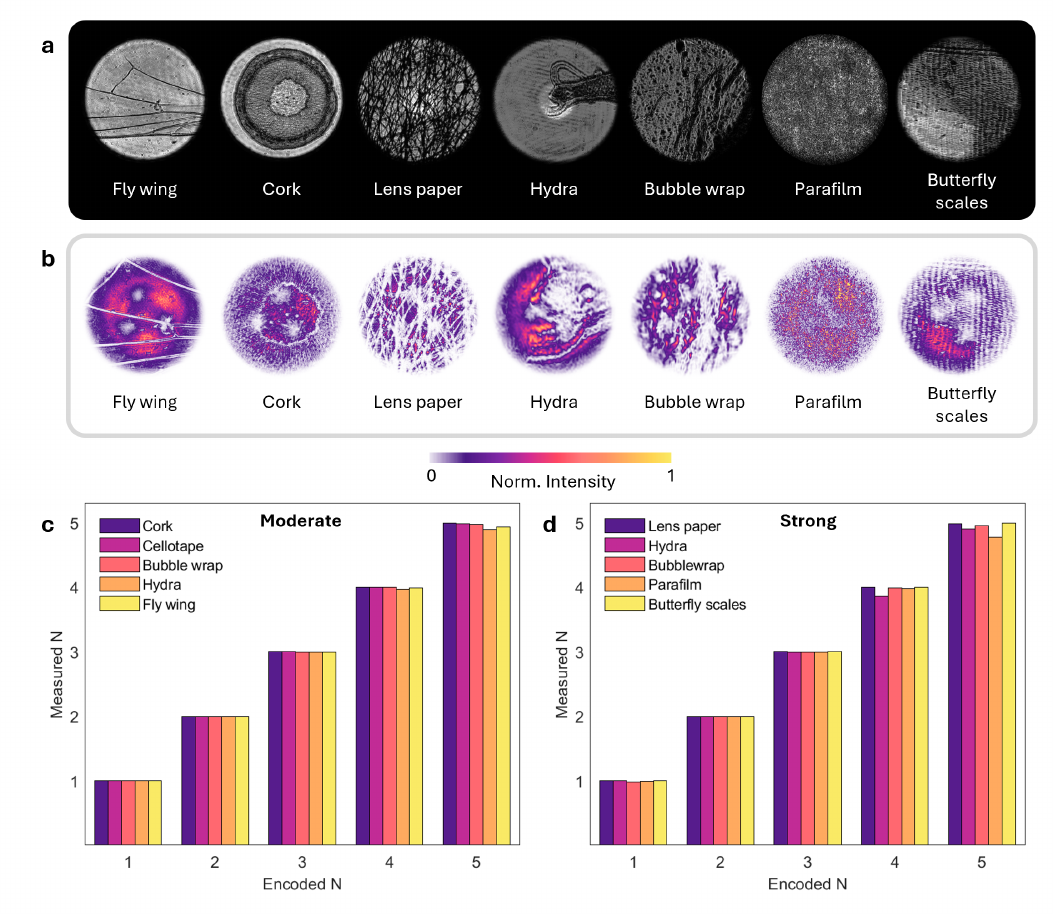}
    \caption{\textbf{Skyrmions through physical random media.} \textbf{a}~Captured images of physical random media including a fly wing, cork, lens paper, cnidarian hydra, bubble wrap, parafilm and butterfly scales. \textbf{b}~Images of the diagonally polarised intensity projection of skyrmion with $N=3$ after passing through each of the samples. \textbf{c}~Plot of the encoded versus measured skyrmion number after passing through physical samples of moderate distorting strength. \textbf{d}~Plot of the encoded versus measured skyrmion number after passing through physical samples of strong distorting strength.}
   \label{fig:realsamples}
\end{figure*}

\noindent \textbf{Inorganic and biological scattering media.} Having completed a controlled and comprehensive test on the topological robustness, we moved on to investigating the robustness of through real-world complex samples. Figure \ref{fig:realsamples} \textbf{a} shows images of the various biological and inorganic random media tested, including a fly wing, cork, lens paper, cnidarian hydra, bubble wrap, parafilm, and butterfly scales. Using the setup described in Figure \ref{fig:setup} \textbf{a}, we generated optical skyrmions and imaged them onto the physical samples, which were then subsequently imaged onto the camera. Figure \ref{fig:realsamples} \textbf{b} shows the diagonally polarised intensity projection for $N=3$ after passing through the various samples. The strength of the induced distortion varied significantly between different samples, with several samples falling into either the weak, moderate or strong distortion regimes. Figure \ref{fig:realsamples} \textbf{c} shows the encoded versus measured skyrmion number for real samples in the moderately distorting regime. The samples used were cork, cello-tape, bubble wrap, cnidarian Hydra, and a fly wing. We see that for all tested topologies and all samples, the measured skyrmion number almost perfectly matches the encoded $N$. Figure \ref{fig:realsamples} \textbf{d} similarly shows results in the strongly distorting regime. The samples used were lens tissue, cnidarian Hydra, bubble wrap, parafilm, and butterfly scales. Different regions of more intense distortion were used for samples that also appeared in Figure \ref{fig:realsamples} \textbf{c}. We see again that for all tested topologies over all samples, the measured $N$ almost perfectly matches the encoded wrapping number. There are some slight deviation for $N=5$, but the measured wrapping number is still close enough to the encoded values as to not incorrectly identify the initial encoded topology. We note that the distortion imparted by the physical samples differs from the digital phase masks as the physical samples do not only distort the phase of the incident light, but also absorbs, blocks, and diffracts light. As such, the lost high spatial frequencies present in the results for the digital phase masks in Figure \ref{fig:resultsdigital} bear a much closer resemblance because of the lost high spatial frequencies which distort the amplitude profile even in the near-field.

\begin{figure*}[t]
    \centering
    \includegraphics[width=\linewidth]{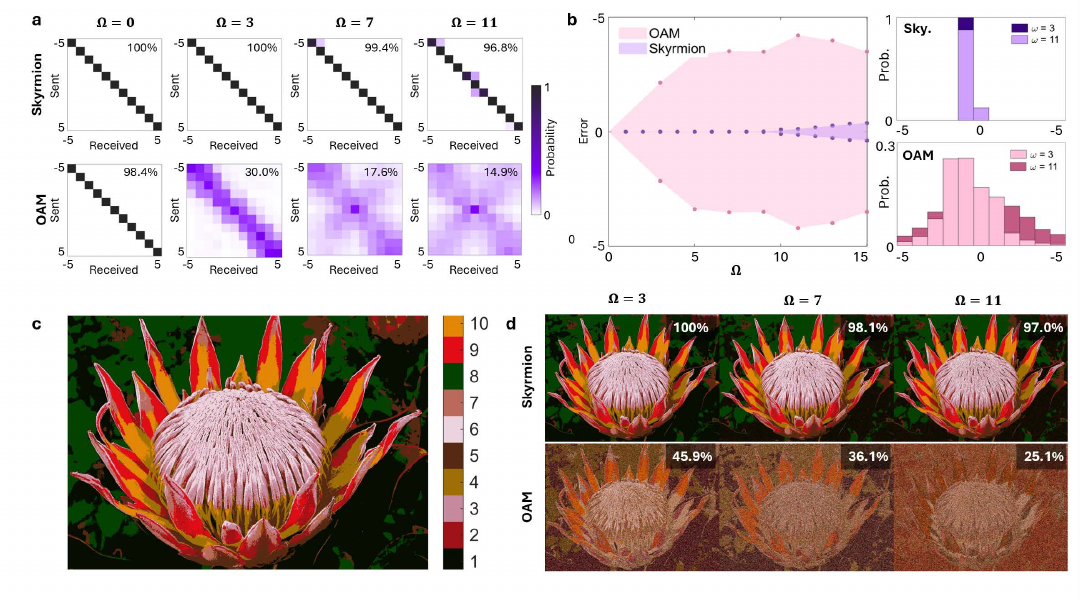}
    \caption{\textbf{Communicating with topological light.} \textbf{a} Crosstalk matrices averaged over 100 random phase masks of skyrmion and OAM beams for various distortion strengths. Inset values indicate the fidelity of the crosstalk matrix. \textbf{b} Comparative spread of error as distortion strength increases for sent skyrmion $N=1$ and OAM mode $l=-1$ with plots on right hand side comparing modal crosstalk for two different distortion strengths. \textbf{c} Encoded 10-colour image of a protea flower. Colourbar shows each colour that is associated with a skyrmion number and an OAM value. \textbf{d} Images transmitted and decoded after passing through digital random masks of various strengths. Top row shows faithful retrieval of the transmitted image even for $\Omega = 11$ with correct pixel colours above 97\% while the bottom row shows the poor performance of OAM  with a visibly distorted transmitted image.    }
    \label{fig:Comms}
\end{figure*}

\noindent \textbf{Communicating through random media.} With the robustness of optical skyrmions firmly established for both simulated and real random media, we move on to compare its stability to that of OAM, which is a commonly employed form of structured light used for encoding and transporting information \cite{willner2015optical}. Figure \ref{fig:Comms} \textbf{a} shows the experimentally measured crosstalk between different skyrmion topologies (top row) as compared to that of OAM (bottom row) through experimentally simulated random media of selected strengths. We see that, in the absence of the perturbation, both forms of light exhibit almost no crosstalk. However, as the strength of the random perturbation increases, the crosstalk between OAM modes gets significantly worse. This is in contrast to the crosstalk between the distinct skyrmion topologies, which remains almost relatively unchanged as $\Omega$ increases. Figure \ref{fig:Comms} \textbf{b} plots a comparison between the error in OAM and skyrmion number through the experimentally simulated random media as the distortion strength increases. We observe that the error in the OAM quickly becomes sufficiently large that it is almost impossible to confidently determine which OAM was initially encoded. In contrast, the error in the topology is sufficiently small throughout the whole range of tested distortion strengths that it is almost always possible to confidently recover the initially encoded skyrmion number. To further illustrate the potential for using topology to transport information through random media, we made use of the image of a protea flower provided in Figure \ref{fig:Comms} \textbf{c}. The image was divided into 10 distinct colours as shown, and each colour was associated with both a skyrmion number and an OAM index. These OAM modes and topological states of light were sent through the experimentally simulated random media. The measured values of the OAM were used to reconstruct the image, shown in the bottom row of Figure \ref{fig:Comms} \textbf{d}. For distortion strengths of $\Omega = 3,7$ and $11$, the reconstructed images have fidelities of 45.9\%, 36.1\% and 25.1\% respectively. When decoding and reconstructing the image using optical skyrmions, shown in the top row of Figure \ref{fig:Comms} \textbf{d}, we see the fidelities of 100\%, 98.1\% and 97\%, almost perfectly matching the initial image and showing a vast improvement over the OAM modes.

\section*{Discussion and conclusion}


\noindent In this work we have provided the first compelling evidence for the robustness of optical skyrmions through real-world random media in the form of inorganic and biological scatterers, and demonstrated its potential for robust information transfer without the need for any corrective measures. Using skyrmions as our example of optical topology, we show almost perfect invariance of the topological wrapping number through channels in regimes where the characteristic length of the perturbation is an order of magnitude smaller than the beam itself, and where very few of the beam's original features can be discerned. A key benefit of our approach is that the invariance of the topology does not depend on knowledge of the medium itself and requires no probe or measurement of the channel for pre- or post-compensation. Unlike their condensed matter cousin, optical topologies have no energy barrier, so the robustness of topology must be shown and not assumed.  Our results are the first to confirm that random and scattering channels can be framed as map preserving for conservation of the wrapping number. This is regardless of the precise geometry of the medium and occurs even when the underlying fields used to construct these topologies are severely aberrated, leaving the topology unchanged.  We have exploited this for the first demonstration of using optical skyrmions to encode and robustly transport information through severely distorting random channels. 

An immediate consequence of this is the robustness of topology through time-varying channels. Our measurement procedure requires a minimum of 4 spatially resolved polarisation intensity measurements which can obtained in a single shot using specially designed projective metasurfaces \cite{li2025disorder} or polarisation sensitive cameras \cite{cox2023real}, with such measurements already demonstrated in rapidly evolving channels such as atmospheric turbulence \cite{peters2023invariance}. 

Intriguingly, while orbital angular momentum is a highly unstable DoF in such channels, the topology upon which it is built is stubbornly resilient.  This work positions topological light as a prime candidate for robust and higher dimensional information encoding with the potential for secure, reliable and resilient communication in noisy and complex environments.  Could it be used for imaging in complex media? We anticipate that topological light may play a role here too, building on the recent trend to replace pixels with spatial modes, both basis elements of ``space'', for imaging, already demonstrated in classical imaging beyond the diffraction limit \cite{paur2016achieving}, in quantum imaging with fewer basis elements \cite{nothlawala2025quantum} and for enhanced spatial sensing \cite{parniak2018beating,rodriguez2022measurement}.  We believe that our work will inspire future research in this direction with the new perspective of topology.

In conclusion, we have successfully demonstrated that topological light is robust to a wide variety of random media. Our results show that it is possible to imbue light with a topological invariant, in our case the skyrmion number, and transport the state through heavily distorting media while still being able to accurately recover the original topological invariant. This approach requires no knowledge of the medium itself, doing away with the need for pre- or post correction. It provides a robust platform with which to encode and transport information through these highly topical channels, showing significant improvement over other forms of structured light such as OAM. We envision the implementation of topological light not only for the classical regime, but also for quantum applications due to the fundamental connection between classical and quantum skyrmions. The potential to reliably see through random media holds immense promise, opening the doorway for robust and reliable classical and quantum communications, imaging, and metrology.

\section*{Competing interests}
The authors declare no competing interests. 

\section*{Acknowledgments}
\noindent C.P. would like to thank Isaac Nape for useful discussions and Bereneice Sephton for valuable feedback regarding this project. C.P. and K.E would like to thank Fazilah Nothlawala for their input on the experimental construction. This work was supported by the South African National Research Foundation/CSIR Rental Pool Programme and the South African Quantum Technology Initiative.

\section*{Author Contributions}
\noindent C.P. and K.E. performed the experiment. K.E. analysed the data. T.K. formulated the binary phase masks and performed numerical simulations. P.O. derived the theory and contributed code for analysing the data. All authors contributed to the writing of the manuscript. A.F. conceived of the idea and supervised the project.

\section*{Code, Data, and Materials Availability}
\noindent Code, data and materials are available upon reasonable request from the corresponding author.

\clearpage
\appendix

\setcounter{section}{0}
\setcounter{figure}{0}
\setcounter{table}{0}
\setcounter{equation}{0}
\setcounter{footnote}{0}
\renewcommand{\thesection}{S\arabic{section}}
\renewcommand{\thefigure}{S\arabic{figure}}
\renewcommand{\thetable}{S\arabic{table}}
\renewcommand{\theequation}{S\arabic{equation}}

\section*{Supplementary: Optical Stokes skyrmion}

\begin{figure}[htpb!] 
    \centering
     \includegraphics[width=0.8\linewidth]{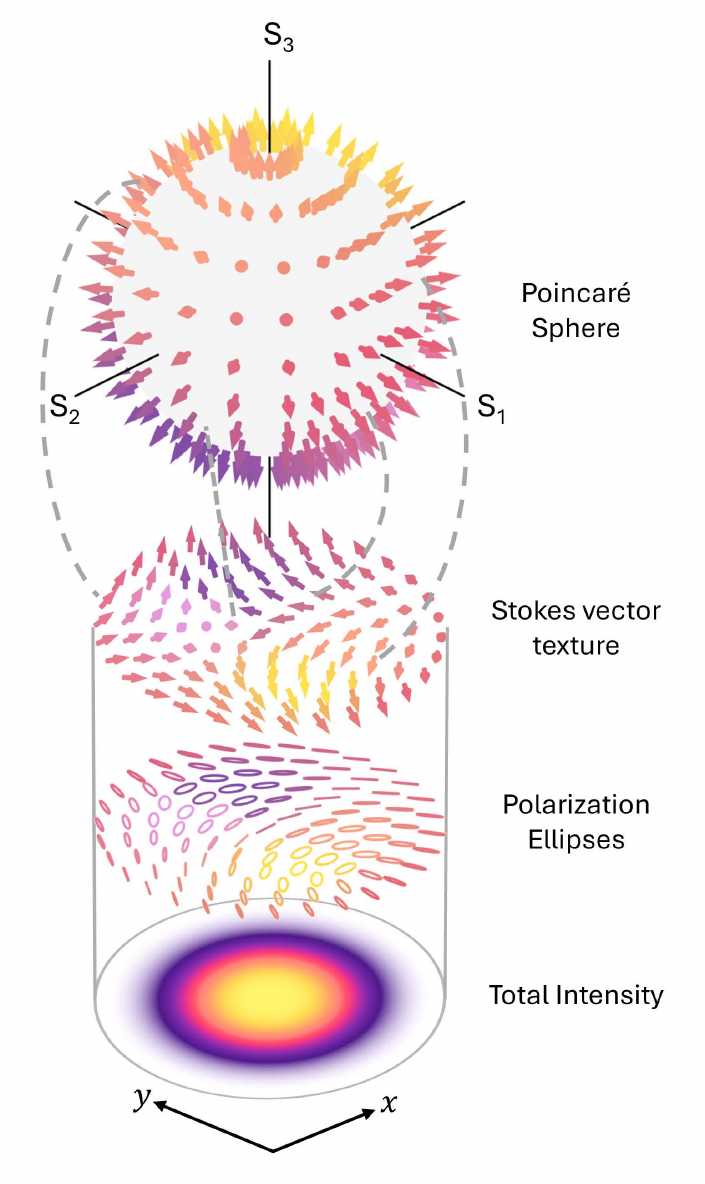}
    \caption{\textbf{Optical Stokes skyrmions.} The combination of two LG modes in two orthogonal polarisation states creates a spatially varying polarisation structure. We may assign a Stokes vector to every point in space, creating a vector textured field. These vectors can be mapped onto the Poincaré sphere, creating a skyrmion topology when $|l_1|\neq|l_2|$. }
    \label{fig:Stokesskyrmion}
\end{figure}

\noindent In this work we consider optical Stokes skyrmions derived from paraxial optical vector fields whose polarisation and spatial degrees of freedom are non-separable \cite{rosales2018review} which can be described using, 
\begin{equation}
\mathbf{U}(\mathbf{r}) = E_1(\mathbf{r})\hat{P}_1 + E_2(\mathbf{r})\hat{P_2}    
\label{eq:General vector beam}
\end{equation}
where $\hat{P}_{1,2}$ represents orthogonal polarisation states and $E_{1,2}(\mathbf{r})$ are Laguerre-Gaussian (LG) spatial modes defined by,
\begin{eqnarray}
    LG_p^l (r, \phi) &=& \sqrt{\frac{2p!}{w^2_0\pi(p+|l|)!}}
          \left(\frac{r\sqrt{2}}{w_0}\right)^{|l|}
          L_p^{|l|}\left(\frac{2r^2}{w^2_0}\right) \nonumber \\
          &\times& e^{-r^2/w^2_0}e^{-il\phi}\,.
          \label{eq:LGdef}
\end{eqnarray} 
Here, $r$ and $\phi$ are the radial and azimuthal coordinates respectively, $w_0$ is the second moment radius of the embedded fundamental Gaussian envelope (typically  setting $w_0 = 0.4$ mm), $p$ is the radial index and $l$ is the azimuthal index that endows the beam with an orbital angular momentum (OAM) of $l\hbar$ per photon. We illustrate this non-separable correlation between position and polarisation in Figure \ref{fig:Stokesskyrmion}, where the spatially varying ellipses denote the polarisation of the electric field at each individual point on the transverse plane. We may quantify the local polarisation state with the Stokes vector, $\mathbf{S}(\mathbf{r})=[S_1\; S_2\; S_3]^T$ whose components are defined by \cite{born2013principles},
\begin{eqnarray}
S_1(\mathbf{r}) &=& E_x^*(\mathbf{r}) E_x(\mathbf{r}) -  E_y^*(\mathbf{r}) E_y(\mathbf{r}) \\ 
S_2(\mathbf{r}) &=& E_x^*(\mathbf{r}) E_y(\mathbf{r}) -  E_y^*(\mathbf{r}) E_x(\mathbf{r}) \\ 
S_3(\mathbf{r}) &=& i\left[E_y^*(\mathbf{r}) E_x(\mathbf{r}) -  E_x^*(\mathbf{r}) E_y(\mathbf{r})\right]. \label{eq:Stokes def}
\end{eqnarray}
Assigning a Stokes vector to every polarisation state in the transverse plane creates a vector textured field as shown in Figure \ref{fig:Stokesskyrmion}. The Stokes vector is normalised such that $\mathbf{S}\cdot\mathbf{S}=1$ to ensure that the vector points along the unit radius Poincaré sphere ($\mathcal{S}^2$) at every point in space. The unit nature of the sphere means every point is uniquely characterised by two angles $(2\psi,2\chi)$ which correspond to the azimuthal and altitude angle, respectively. The Stokes vector has a unique orientation for every possible polarisation state, meaning we can map every point on the transverse plane to a point on the Poincaré sphere, defining a map from $\mathcal{R}^2$ to $\mathcal{S}^2$. When a vectorial state of light of the form given by Equation \ref{eq:General vector beam} has $|l_1| \neq |l_2|$, the field contains every possible relative phase and relative amplitude between the two components. It must then contain every possible polarisation state at least once. The number of times every polarisation state appears dictates how many times one wraps the Poincaré sphere. A skyrmion topology exists when one wraps the Poincaré sphere an integer number of times. (i.e how many times $\mathcal{R}^2$ wraps $\mathcal{S}^2$) \cite{Gao2020paraxial}. The wrapping number $N$ uniquely characterizes the topology and can be calculated from the normalized Stokes parameters as follows,
\begin{equation}
    N =  \frac{1}{4\pi} \int_{\mathcal{R}^2} \left( \textbf{S} \cdot \frac{\partial\textbf{S}}{\partial x} \times \frac{\partial\textbf{S}}{\partial y}\right) \text{d}x \text{d}y 
    \label{eq:skyrmion Surface}
\end{equation}
which simplifies down for states in form of Equation \ref{eq:General vector beam} to,
\begin{equation}\label{eq:SkrymeShortcut}
    N = n |l_1-l_2| \,,
\end{equation}
where $n = -1$ if $l_2 > l_1$ and $n=1$ if $l_2 < l_1$. 

\section*{Supplementary: Experimental determination of the skyrmion wrapping number} \label{sec:experimental_determination}

\begin{figure*}[htpb!]
    \centering
    \includegraphics[width=0.85\linewidth]{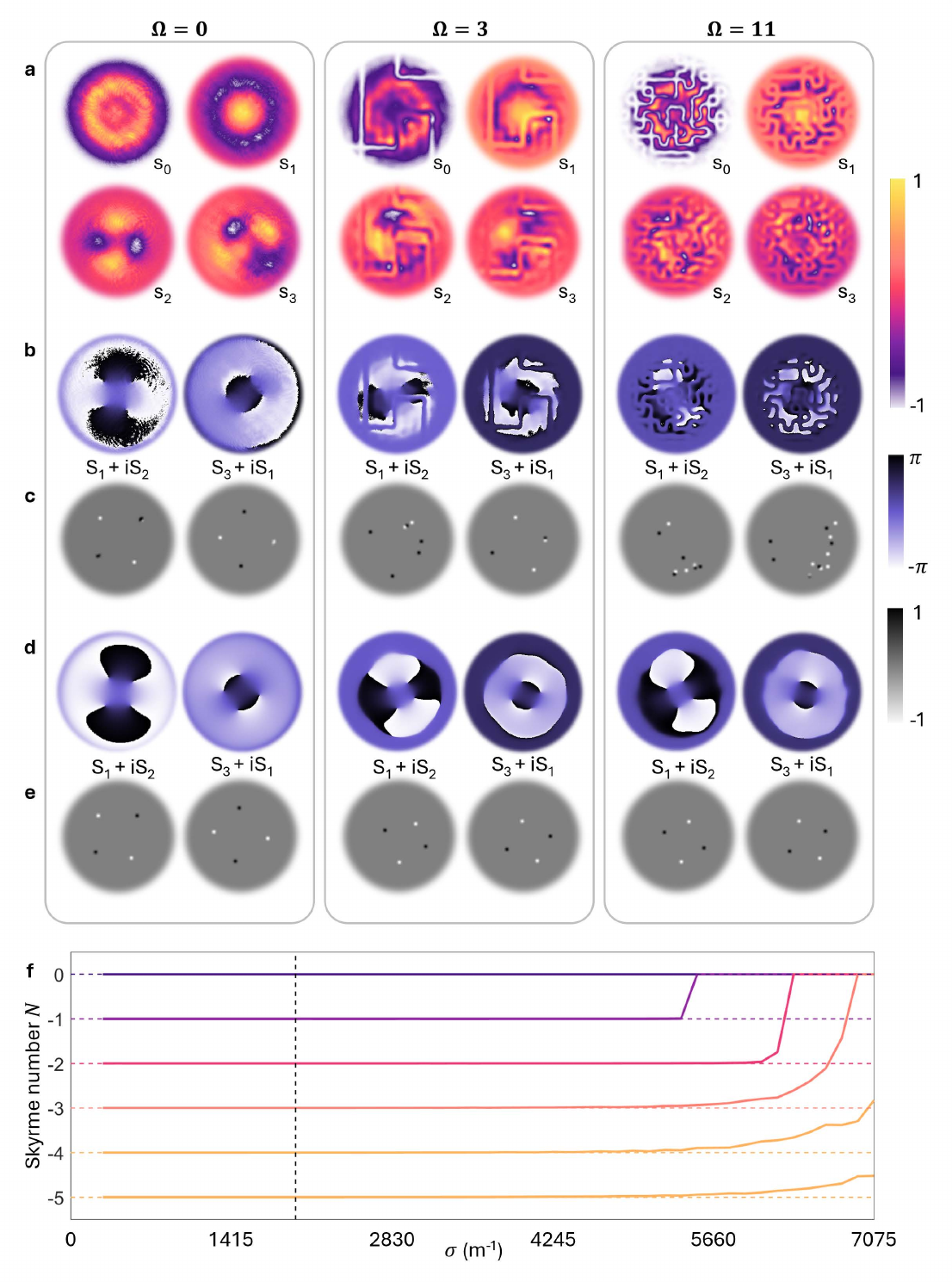}
    \caption{\textbf{Measuring polarisation singularities.} \textbf{a} Examples of experimentally obtained Stokes parameters for a skyrmion of $N=2$ ($l_1 = 2$ and $l_2 = 0$) for distortion strengths $\Omega = 0,3$ and $11$. \textbf{b} The polarisation phases the fields $P_2 = S_3 + iS_1$ and $P_3 = S_3 + iS_1$ calculated from the experimentally obtained Stokes parameters. \textbf{c} The calculated circulation giving the location and charges of the singularities for the fields $P_2 = S_3 + iS_1$ and $P_3 = S_3 + iS_1$ \textbf{d} the polarisation phase of the fields $P_2 = S_3 + iS_1$ and $P_3 = S_3 + iS_1$ after the application of a Gaussian filter with a kernel size $\sigma = 1981$~m$^{-1}$. \textbf{e} The calculated circulation of the filtered experimental results. \textbf{f} The measured skyrmion number of simulated skyrmions $N=0,-1,-2,-3,-4$ and $-5$ plotted against the size of the kernel of the applied Gaussian filter $\sigma$. The vertical black dotted line indicates the size of the kernel used to process the experimental results. }
    \label{fig:ExpStokes}
\end{figure*}

\noindent \textbf{Topological characterisation of skyrmions.} To experimentally determine the skyrmion number, $N$, of an experimentally generated field, one must determine the wrapping number, typically calculated according to Equation \ref{eq:skyrmion Surface}. In this work, we use the approach proposed by McWilliam \textit{et al.} \cite{McWilliam2023topological}, which makes use of a contour integral over complex polarisation fields. Primarily, we make use of their result which expresses the skyrmion number as follows,
\begin{equation} \label{eq:line_int}
     N = \frac{1}{2} \left( \sum_{j} S_{z}^{(j)} N_{j} - \bar{S_{Z}^{\infty}} N_{\infty} \right) \,,
\end{equation}
where $N_j$ is the charge of individual phase singularity at position $j$ in the field $S_x + iS_y $,  $S_z^{(j)}$ is the value of the Stokes parameter $S_z$ at the point $j$, $N_{\infty}$ is the result of the contour integral at infinity and $S_z^{(\infty)}$ is the value of the Stokes parameter $S_z$ as $r\rightarrow \infty$. Any of the Stokes parameters ($S_1$, $S_2$ and $S_3$) can take the place of $S_z$, with the other two ordered taking the place of $S_x$ and $S_y$.\\

\noindent \textbf{Stokes Polarimetry.} The Stokes parameters of the generated optical skyrmions were experimentally measured using six polarisation intensity projections,
\begin{eqnarray} \label{eq:stokesS0}
    s_{0} &=& I_{H} + I_{V} \\
\label{eq:stokesS1} 
    s_{1} &=& I_{H} - I_{V}\\
\label{eq:stokesS2}
    s_{2} &=& I_{D} - I_{A}\\
\label{eq:stokesS3}
    s_{3} &=& I_{R} - I_{L} \,.
\end{eqnarray}
where the subscripts H, V, D, A, R and L represent horizontal, vertical, diagonal, antidiagonal, right circular and left circular polarisations respectively. A linear polariser and half-wave plate was used to acquire the intensity projections for the linear projections and a linear polariser in combination with a quarter-wave plate  and half-wave plate for the circular projections \cite{singh2020digital}. Experimentally obtained Stokes parameters $s_j$ are shown in Figure \ref{fig:ExpStokes} \textbf{a} for $N=2$ ($l_1 = 2$ and $l_2 = 0$) for distortion strengths $\Omega =0,\,3$ and $11$. Equation \ref{eq:line_int} requires the locally normalised Stokes parameters $S_j$ which were computed from the experimentally obtained Stokes parameters $s_j$ according to,
\begin{equation}
    S_j =\frac{s_j}{\sqrt{s_1^2+s_2^2+s_3^2}}
\end{equation}
\\

\noindent \textbf{Polarisation singularities.} In order to determine the location and charge of the singularities $N_j$, the polarisation fields $P_j$ were calculated using,
\begin{eqnarray} \label{eq:polphase1}
P_{1} &=& S_{2} + iS_{3}\\
\label{eq:polphase2}
P_{2} &=& S_{3} + iS_{1}\\
\label{eq:polphase3}
P_{3} &=& S_{1} + iS_{2}\,.
\end{eqnarray}
Example phases of the complex fields $P_2$ and $P_3$ are plotted in Figure \ref{fig:ExpStokes} \textbf{b} for $N=2$ ($l_{1,2} = 2,01$) for distortion strengths $\Omega =0,\,3$ and $11$, calculated from experimental measurements. $P_1$ was excluded as it was not used in the calculation of the skyrmion number. Once these fields were calculated, the locations and charges of the individual singularities were computed using a numerical equivalent to the curl $\nabla \times$ operation proposed in Ref \cite{chen2007detection} termed the circulation $D$. The circulation is defined as,
\begin{eqnarray}
    D^{m,n} = &\frac{d}{2}& ( G_x^{m,n} + G_x^{m,n+1} +G_y^{m,n+1} + G_y^{m+1,n+1} \nonumber \\ &-  &G_x^{m+1,n+1} - G_x^{m+1,n} -G_y^{m+1,n} \nonumber \\
    &-& G_y^{m,n})\,.
\end{eqnarray} 
Here, $D^{m,n}$ represents the value of the circulation of the pixel in the $n$-th row and $m$-th column. $G_x^{m,n}$ and $G_y^{m,n}$ are the phase gradient in the horizontal and vertical direction of the pixel in the $n$-th row and $m$-th column, respectively and $d$ is the pixel size. Typically, the circulation will return a 0 value if there is no singularity at that pixel and a nonzero value if there is. The magnitude of the circulation indicates the charge of the singularity and the sign indicates the direction/handedness of the singularity. These values were then substituted into Equation \ref{eq:line_int} to calculate the skyrmion number. Examples of the circulation calculated using $P_2$ and $P_3$ are shown in Figure \ref{fig:ExpStokes} \textbf{c} for a skyrmion $N=2$ ($l_{1,2} = 2,0$) for distortion strengths $\Omega =0,\,3$ and $11$. It is clear that as the distortion strength increases, more and more singularities are detected which will lead to a measured skyrmion number that differs from the encoded one. These additional singularities may occur due to several reasons, including ambient environmental noise, shot noise of the detector, or from the amplitude distortions exhibited by the beam. However, we devised a post-processing procedure that allowed for the recovery of the original skyrmions number that is agnostic to structure of the original beam. \\

 \noindent \textbf{Post-processing of experimental data.} The first step was the application of a low-pass 2D Gaussian filter with smoothing kernel with standard deviation given by $\sigma=1981$~m$^{-1}$. The size of the smoothing kernel was kept constant for all $N$ and $\Omega$ tested, ensuring an unbiased procedure to remove noise and high frequency contributions to final calculation. Examples of polarisation phases $P_2$ and $P_3$ after the application of the Gaussian filter to the measured intensity projections is shown \ref{fig:ExpStokes} \textbf{d} for a skyrmion with $N=2$ ($l_1 = 2$ and $l_2 = 0$) for distortion strengths $\Omega =0,\,3$ and $11$. The circulation calculated from the filtered experimental results is shown in Figure \ref{fig:ExpStokes} \textbf{e}. The choice of filter strength was informed by numerical simulation. Simulated skyrmions with $N=0,-1,-2,-3,-4$ and $-5$ and the exact parameters of the experimentally generated beams were exposed to Gaussian filters of increasing kernel size, from $\sigma = 0$ to $\sigma = 7075~m^{-1}$. The skyrmion number was calculated after the application of each filter strength, with the results shown in \ref{fig:ExpStokes} \textbf{f}. We see that all the skyrmion numbers remain unchanged under the action of the Gaussian filter until the kernel reaches a size of $\approx 5300$~m$^{-1}$. This provides an upper limit on the allowable range of kernel sizes that could be applied to the experimental data. The choice of $\sigma=1981$~m$^{-1}$ was the lowest filter strength within this range that allowed for consistent agreement of the measured skyrmion number with the encoded value. In addition to Gaussian filtering, we also implemented a standard intensity threshold, only counting singularities that were found in regions with intensities above the noise floor. To determine the noise floor, regions near the edges of the captured CCD images, far removed from the generated beam, were isolated and the intensity values averaged. The average noise value over multiple measurements was found to be $\approx 2\%$ of the maximum intensity. Therefore, any singularity detected in regions with intensities lower than 2\% of the maximum beam signal were disregarded from the calculation of the skyrmion number. The same post-processing parameters and procedure were applied irrespective of the inputted topological number, the distortion strength of the medium and for both the digital phase masks and the physical samples. 

\section*{Supplementary: Formulation of digital random phase masks} \label{sec:supp_formulation}

\begin{figure*}[htbp!] 
    \centering
     \includegraphics[width=0.8\linewidth]{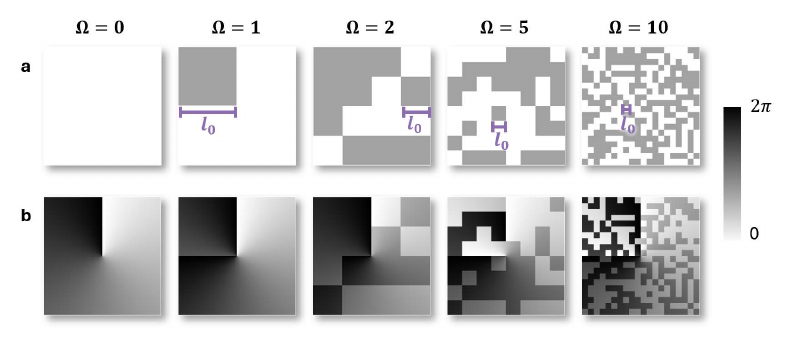}
    \caption{\textbf{Binary random phase masks.} \textbf{a} Example binary phase masks for various distortion strengths, where the phase of the superpixels alternates between 0 and $\pi$. The size of each superpixel is equal to the correlation length $l_0$. \textbf{b} The effect of the binary phase masks on the phase of an LG beam with $l=1$.  }
    \label{fig:binary_masks}
\end{figure*}

 \noindent The mathematical description of scattering, complex, and random media varies greatly depending on the type of medium studied. As a result, it is challenging to make general statements about such media. Fortunately, in the case of transverse structured light fields, there exists a universal crosstalk law that describes the coupling between spatial modes induced by any perturbing channel that depends solely on the beam's spatial size $w$ and the correlation length of the medium $l_0$ \cite{bachmann2024universal}. Building on this framework, we may characterise the effects of an arbitrary scattering medium with the use of binary phase masks $e^{i\Theta(\mathbf{r})}$, where the applied phase $\Theta(\mathbf{r})$ alternates between the values of 0 and $\pi$ in distinct regions across the transverse plane. The size of these regions corresponds to the desired correlation length of the medium $l_0$. A screen with larger regions of constant phase results in an overall more uniform phase profile that will exhibit a minimal perturbing effect on the beam. A screen with smaller regions of constant phase results in a more rapidly varying phase profile and consequently a more severe perturbing effect on the beam.

Digital devices such as a spatial light modulator (SLM) provide the perfect toolkit to implement this approach, allowing for the on-demand tailoring of the applied perturbing phase screen. We applied the binary phase masks as part of the mode generation step, where the complex field described by the $LG_l^p$ function was computed and then multiplied by the binary phase masks. This combined complex field was then used to compute the final digital hologram according to Ref \cite{arrizon2007pixelated}. 

Different values of $l_1$ and $l_2$ correspond to LG beams of different second moment radius $w_{|l|}$ (where $w_{|l|}=w_0\sqrt{1 + |l|}$ and $w_0$ is the second moment radius of the embedded fundamental Gaussian beam). To adequately apply the universal crosstalk law, both the beam's spatial extent and the correlation length of the medium $l_0$ must be accounted for when characterizing the distortion strength of the channel. We therefore defined the unitless parameter,
\begin{equation}
    \Omega = 2w_{|l_{max}|}/l_0\,,
    \label{eq:Dist strength}
\end{equation}
where $w_{|l_{max}|}$ corresponds to the second moment of the beam in the vectorial combination with the larger OAM index. This ensured that the increased transverse spatial size of higher order LG modes did not disproportionately impact the robustness of higher order topologies.

With a chosen value of $\Omega$, the region of interest was then divided in super-pixels whose side-length was equal to $l_0$. We illustrate this in Figure \ref{fig:binary_masks} \textbf{a}, where example masks are shown for distortion strengths $\Omega = 0,1,2,5$ and $10$. We chose a square super-pixel geometry for the phase masks as to easily match the geometry of the underlying pixels of the SLM however, alternative geometries could have been used. Each super-pixel was then randomly assigned a phase of $0$ or $\pi$, shown as white for 0 and gray for $\pi$ in figure \ref{fig:binary_masks} \textbf{a} in before being applied to the complex $LG_l^p$ field. The effect of the binary masks on the phase of the LG beam is shown in Figure \ref{fig:binary_masks} \textbf{b}. We see that the higher the distortion strength, the more variation there is in the beam's phase profile and thus the more aberrated the phase becomes. We made use of the built-in random number generation capability of MATLAB to generate multiple random realisations for phase masks with the same $l_0$, allowing us to build up the necessary statistics to accurately emulate the desired universal crosstalk law.

\section*{Supplementary: Characterisation of the distortion strength of digital masks, inorganic and biological samples} 

\begin{figure*}[htpb!]
    \centering
     \includegraphics[width=0.9\linewidth]{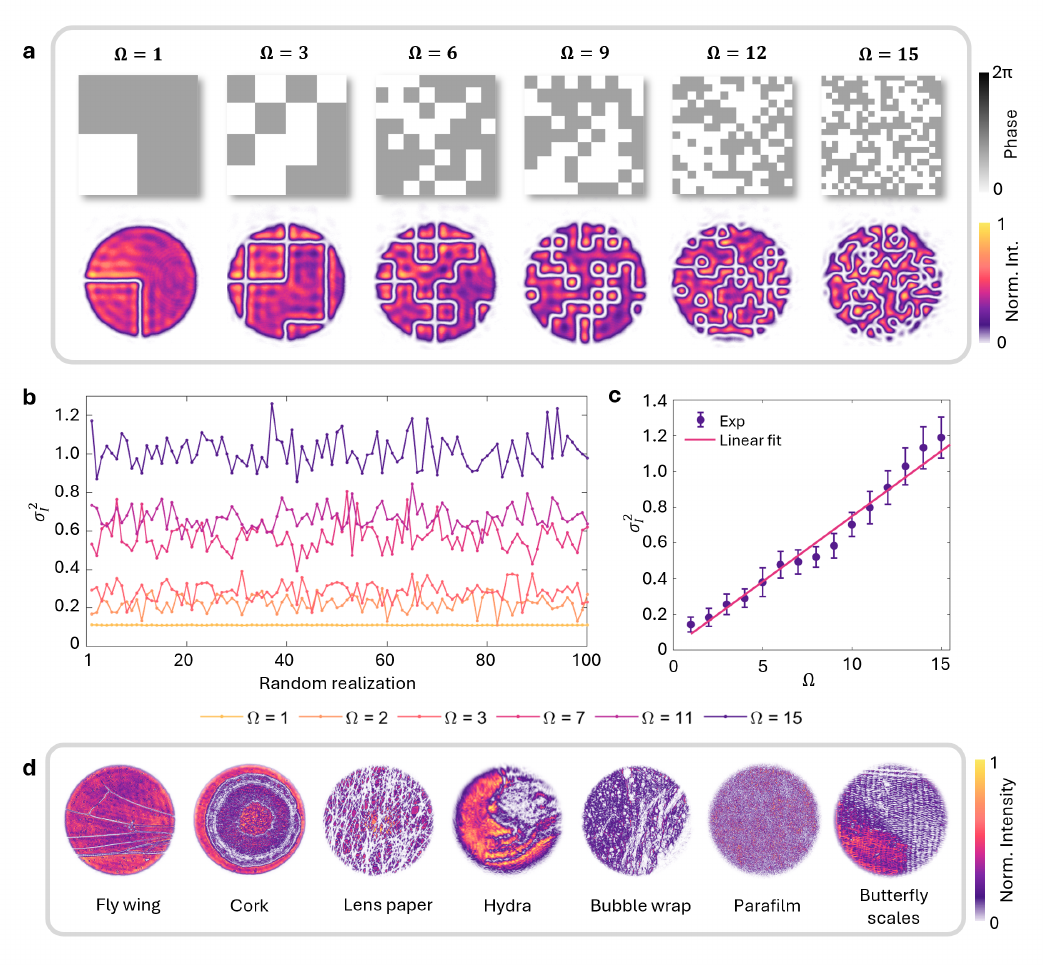}
    \caption{\textbf{Scintillation index and distortion strength.} \textbf{a} Binary phase masks and the corresponding experimental images of an aberrated circular beam of uniform phase and intensity used to compute the scintillation index. \textbf{b} The measured scintillation index for 100 random realisations of binary phase screens for various selected distortion strengths. \textbf{c} Experimentally measured correspondence between the encoded distortion strengths and measured scintillation index of the digital phase masks. We see a linear relationship between the two quantities, with $R^2=0.9802$. \textbf{d} Experimental images of an aberrated circular beam of uniform phase and intensity after passing through selected physical samples used to compute the scintillation index. }
    \label{fig:ScintIndex}
\end{figure*}

 \noindent A versatile digital toolkit allows for the precise control over both beam generation and the imparted phase distortion. However, some of this control is lost when dealing with real-world random samples whose correlation length is fixed. In order to quantitatively compare digital experimental results with results obtained using physical samples, an intermediate measure was needed. We opted to use the scintillation index $\sigma_I^2$ which is commonly used in studies of atmospheric turbulence and quantifies the variance in the irradiance fluctuations of a plane wave that has propagated through a perturbing channel \cite{peters2025structured}. It is easily obtainable in experiment as it requires intensity only measurements of the perturbed beam which are easily obtained with the CCD. Mathematically it is given by,
\begin{equation} \label{eq:scint_formulae}
    \sigma_I^2 = \frac{\langle I^2(x,y)\rangle}{\langle I(x,y)\rangle^2}-1 \,,
\end{equation}
where $I(x,y)$ represents the spatially varying intensity profile of the wave and $\langle \cdot \rangle$ represents the average over the transverse plane.

To investigate the relationship between the distortion strength $\Omega$ and scintillation index $\sigma_I^2$ using the digital random phase masks, we made use of the complex amplitude modulation capabilities of the SLM to produce a circular beam of uniform phase and constant intensity with set diameter $D$, shown in Figure \ref{fig:ScintIndex} \textbf{a}, where examples of random binary phase masks of various strengths (top row) were applied to circular beams of uniform phase and intensity (bottom row). The phase of this beam was then modulated with binary random phase masks of varying correlation lengths $l_0$ to achieve a range of distortion strengths $\Omega = D/l_0$. The measurements were performed in the same configuration as shown in Figure 2 \textbf{a} of the main text, with the SLM plane being imaged onto the CCD. The binary phase masks are a phase only perturbation applied using the SLM. However, we observe at the detector both phase and amplitude distortions. The phase distortions manifest as expected but the amplitude distortions are a result of the phase perturbation and the limitations of the imaging systems used. The phase perturbation consists of binary phases, with abrupt phase jumps between 0 and $\pi$ and thus have very high spatial frequencies in their Fourier decomposition. A 4f imaging system like the one used in this experiment is normally capable of perfectly reconstructing the phase and amplitude profile of any incident field. However, due to the high frequencies present in the binary phase masks, some of the light in the perturbed beam is not adequately captured by the lenses, as it diffracts and spreads out rapidly in propagation. We therefore observe lines of missing intensity in the image plane of the SLM, corresponding to the boundary between the phase jumps and where the high frequency components were unable to adequately propagate through the imaging system \cite{saleh2019fundamentals4}. This is clearly seen in Figure \ref{fig:ScintIndex} \textbf{a} where we see that the sharp lines of missing intensity occur in the exact same locations as the phase jumps in the the phase masks. It is also clear that the degree of the amplitude distortion is proportional to the the number of phase jumps in the binary phase screen and thus the encoded distortion strength $\Omega$. We therefore used the scintillation index $\sigma_I^2$ to experimentally quantify the distortion strength and compare it to the measured scintillation index of the physical sample, which also distorts the intensity of an incident beam in a similar manner.

Figure \ref{fig:ScintIndex} \textbf{b} shows the measured scintillation index over 100 random realisations of the binary phase screens for various encoded distortion strengths. Although the scintillation index fluctuates between different realisations of masks of the same distortion strength, we can clearly see that higher $\Omega$ corresponds to a larger scintillation index $\sigma_I^2$. Figure \ref{fig:ScintIndex} \textbf{c} shows the averaged results of the scintillation index $\sigma_I^2$ as a function of the distortion strength $\Omega$. Each data point represents the average over 100 realisations, with the error bars indicating the standard deviation. As expected, we observe a positive, linear relationship between the scintillation index $\sigma_I^2$ and the distortion strength $\Omega$. This plot then allows us to infer the distortion strengths of the physical samples to a sufficient degree of precision to ensure the distortion induced by the samples is similar in strength to that of the binary phase masks and class the samples into similar regimes of weak, moderate and strong distortion. Specifically we defined weak distortion as $\sigma_I^2<0.4$, moderate: $0.4 \le \sigma_I^2 \le 0.6$, and strong: $\sigma_I^2>0.6$. The scintillation indices for the physical samples were measured in a manner similar to that of the digital phase masks. Regions of the samples were probed by the uniform circular beam profiles, and the intensity of this beam was captured by the CCD. The scintillation index was then calculated and that region of the sample was then grouped into one of the three defined regimes. Some samples had regions of varying distortion strength, and so were subsequently grouped into different regimes according to which region of the samples was being imaged through. Images of samples with the uniform intensity probe beam are shown in \ref{fig:ScintIndex} \textbf{d}. Table \ref{tab:scint} shows the measured values of the scintillation index for various physical samples used in this work.

\begin{table*}[htbp!]
\centering
\begin{tabular}{llllll}\toprule 
\multicolumn{2}{c}{\textbf{Weak }} & \multicolumn{2}{c}{\textbf{Moderate }} & \multicolumn{2}{c}{\textbf{Strong }} \\ 
\multicolumn{2}{c}{\textbf{$\sigma_I^2<0.4$}} & \multicolumn{2}{c}{\textbf{$0.4\le \sigma_I^2 \le 0.6$}} & \multicolumn{2}{c}{$\sigma_I^2>0.6$} \\ \hline

\multicolumn{1}{l}{Dirty glass}       & 0.19               & \multicolumn{1}{l}{Cork (2)}                  & 0.45                    & \multicolumn{1}{l}{Lens paper}                 & 1.34              \\
\multicolumn{1}{l}{Rana ova}                & 0.20                & \multicolumn{1}{l}{Cellotape (2)}             & 0.48                    & \multicolumn{1}{l}{Hydra (3)}           & 0.66              \\
\multicolumn{1}{l}{Cellotape}               & 0.27               & \multicolumn{1}{l}{Bubblewrap (2)}            & 0.44                    & \multicolumn{1}{l}{Bubblewrap (3)}      & 0.74              \\
\multicolumn{1}{l}{Bubblewrap}              & 0.39               & \multicolumn{1}{l}{Hydra (2)}                 & 0.43                    & \multicolumn{1}{l}{Parafilm}                   & 1.16              \\
\multicolumn{1}{l}{Fly wing}                & 0.24               & \multicolumn{1}{l}{}                                 &                         & \multicolumn{1}{l}{Butterfly scales}           & 0.68              \\
\multicolumn{1}{l}{Cork}                    & 0.28               & \multicolumn{1}{l}{}                                 &                         & \multicolumn{1}{l}{}                           &                   \\
\multicolumn{1}{l}{Hydra}                   & 0.35               & \multicolumn{1}{l}{}                                 &                         & \multicolumn{1}{l}{}                           &                  \\
\botrule
\end{tabular}
\caption{\textbf{Measured scintillation index for physical samples.} Each distortion regime was computed using Equation \ref{eq:scint_formulae} for a plane wave incident on the sample. Values in brackets indicate different regions of interest used on the same sample which distorted the intensity profile of the plane wave beams to different extents and thus had different scintillation indices.} 
    \label{tab:scint}
\end{table*}

\section*{Supplementary: Additional Experimental data}

 \noindent \textbf{Experimental skyrmion Beams.} Eleven different skyrmion numbers were tested verifying the reliable generation and measurement of the topology in the absence of binary phase masks ($\Omega = 0$). The results, shown with the OAM combination used to generate the topologies, are shown in table \ref{tab:ideal_skrymenumbers} for data averaged across 100 mask realisations. We see that the averaged values are very close to the encoded topology values which were inferred by the difference in OAM as described by Equation \ref{eq:SkrymeShortcut}. The standard deviation was also several orders of magnitude smaller than the measured values indicating a stable topology generating system and reliable measurement of the skyrmion number for the generated states. It is worth noting that for $N=0$, the standard deviation is perfectly zero. This is because no singularities are detected for this case, meaning the skyrmion number is exactly 0 for each measurement, and did not fluctuate over the measurement period.\\
 
\begin{table}[htbp!]
    \centering
    \begin{tabular}{cccc} \toprule
       \boldmath{ $\ell_1 $ } & \boldmath{ $\ell_2$ }&\boldmath{ N$_{th}$ }& \boldmath{N$_{exp}$ ($\Omega = 0$)}\\\hline
        -2 & 3 & -5 & $-4.99363\pm 0.00200$ \\
        -1 & 3 & -4 & $-3.99577\pm 0.00115$ \\
        0 & 3 & -3 & $-2.99027\pm  0.00343$ \\
        0 & 2 & -2 & $ -1.99778\pm 0.00083$ \\
        0 & 1 & -1 & $ -0.99873\pm 0.00128$ \\
        0 & 0 & 0 & $0 \pm 0$\\ 
        1 & 0 & 1 & $0.99535 \pm 0.00256 $\\
        2 & 0 & 2 & $1.99757 \pm 0.00078$ \\
        3 & 0 & 3 & $2.99155\pm 0.00252$ \\
        1 & -3 & 4 & $3.99547 \pm 0.00134$ \\
        2 & -3 & 5 & $4.99318\pm 0.00207$ \\ \botrule 
    \end{tabular}
    
    \caption{\textbf{Unperturbed experimental skyrmion numbers.} Evaluated skyrmion numbers $N_{exp}$ after post-processing averaged over 100 images from the polarisation field $S_1+iS_2$ for the ideal case with no added perturbation or distortion ($\Omega = 0$) applied demonstrating reliable generation and measurement of optical skyrmions. The indices $l_1$ and $l_2$ indicate the OAM of the component LG modes. $N_{th}$ is the theoretically computed skyrmion number using Equation \ref{eq:SkrymeShortcut}. } 
    \label{tab:ideal_skrymenumbers}
    
\end{table}

\begin{figure*}[h]
    \centering
     \includegraphics[width=\linewidth]{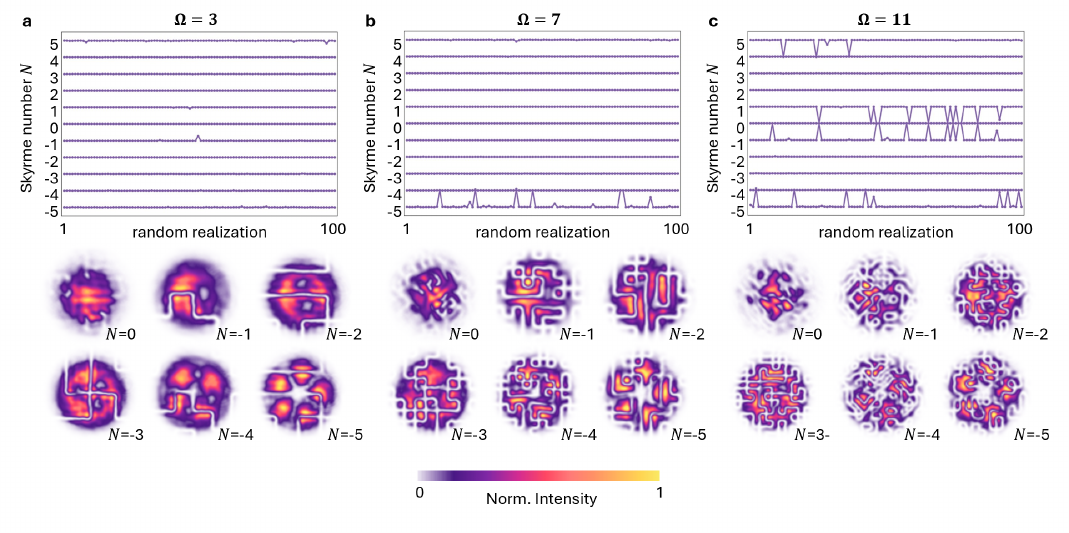}
    \caption{\textbf{Robustness through digital random phase masks.} The measured skyrmion numbers for each random realisation of the binary phase masks for encoded skyrmion numbers $N=-5$ to $N=5$ for strengths \textbf{a} $\Omega = 3$, \textbf{b} $\Omega = 7$ and \textbf{c} $\Omega = 11$ along with the diagonal intensity projections of selected random realisations for each of the encoded skyrmion numbers for each strength.  }
        \label{fig:SeedExpResults}
\end{figure*}

 \noindent \textbf{Digital phase masks.} Figure \ref{fig:SeedExpResults} shows the measured skyrmion number for each random realisation of the binary phase masks for all 11 of the tested skyrmion numbers for distortions strengths $\Omega =3,\,7 $ and $11$, along with examples of the diagonal polarisation intensity projection of the perturbed beam for $N = 0,-1,-2,-3,-4$ and $-5$ for these strengths. These plots correspond to the data shown in Figure 3 of the main text. Overall, we observe that the measured skyrmion number remains almost unchanged through the random realisations for most $N$ over all shown $\Omega$. This shows not only why the average skyrmion number remains close to the encoded value, but also why the standard deviations measured were small, and in some cases negligible, in relation to the measured $N$. This demonstrates that the resilience of the optical skyrmion is independent of the exact structure of the simulated medium and implies one does not need to measure or probe the medium beforehand to ensure the skyrmion number remains unchanged. We do see some variation in the higher order topologies of $N=\pm5$, this can possibly be attributed to the complex polarisation and singularity structure of these higher order topologies, but further investigation is required to confirm this.\\

\begin{figure*}[htbp!]
    \centering
     \includegraphics[width=\linewidth]{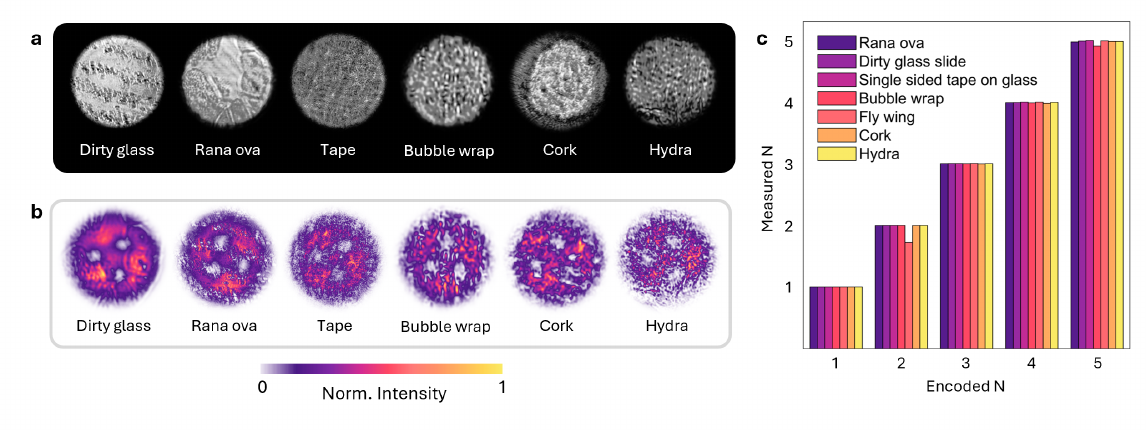}
    \caption{\textbf{Weak distorting samples.} \textbf{a} Images of the seven samples used to test the robustness of optical skyrmions in the weak distorting regime. \textbf{b} The antidiagonal intensity projection of $N=3$ after passing through the physical samples. \textbf{c} Graph showing the encoded vs measured $N$ for the weak distorting samples. }
        \label{fig:BioSamplesWM}
\end{figure*}

\noindent \textbf{Weakly distorting physical samples.} In addition to the moderate and strong distorting physical samples, we also tested physical samples whose distortion fell into the weak regime. We show images of these samples in Figure \ref{fig:BioSamplesWM} \textbf{a} that include a dirty glass slide, rana ova, transparent tape, bubble wrap, cork and a cnidarian Hydra. Some of these samples were used in the moderate and strong regimes, but here the beam size was reduced to decrease the distortion strength according to Equation \ref{eq:Dist strength}. Figure \ref{fig:BioSamplesWM} \textbf{b} shows the antidiagonal polarisation projection intensity of a skyrmion $N=3$ after passing through the samples. We notice fluctuations in the beam profiles, but are still able to perceive noticeable features of the beam, including the three regions of null intensity which correspond to the presence of phase singularities. The presence of these singularities directly correlates to positions of the polarisations singularities and therefore how challenging it would be to calculate the skyrmion number after the distortion. Figure \ref{fig:BioSamplesWM} \textbf{c} shows the encoded vs measured $N$ for the weakly distorting physical samples. We see almost perfect agreement and thus the preservation of the topology in all 5 of the tested skyrmion numbers and across all the 7 samples tested.\\

\section*{Supplementary: Communicating using topological light}

 \noindent To demonstrate the advantages of using topological light, we performed an experiment comparing the robustness of optical skyrmions and OAM beams by using each to encode and detect information sent through our digital random masks. OAM was chosen as it is the most common form of transverse structured light used for optical communication \cite{willner2015optical}. To perform this experiment, we used the same experimental setup shown in Figure 2 \textbf{a} of the main text which is capable of generating vectorial combinations of scalar beams. Information was encoded using the discrete set of wrapping numbers available using skyrmion beams and in the discrete set of OAM values of LG beams. After perturbation by random phase masks, the wrapping number of the skyrmion beams and the OAM of the perturbed LG beams was measured and used to decode the transmitted image shown in Figure 5 \textbf{c} of the main text. In this section we provide details of the OAM measurement process, how the crosstalk and error of the skyrmions and LG beams were quantified, and the scheme used to encode and decode the image.\\

\begin{figure*}[t!]
    \centering
     \includegraphics[width=0.7\linewidth]{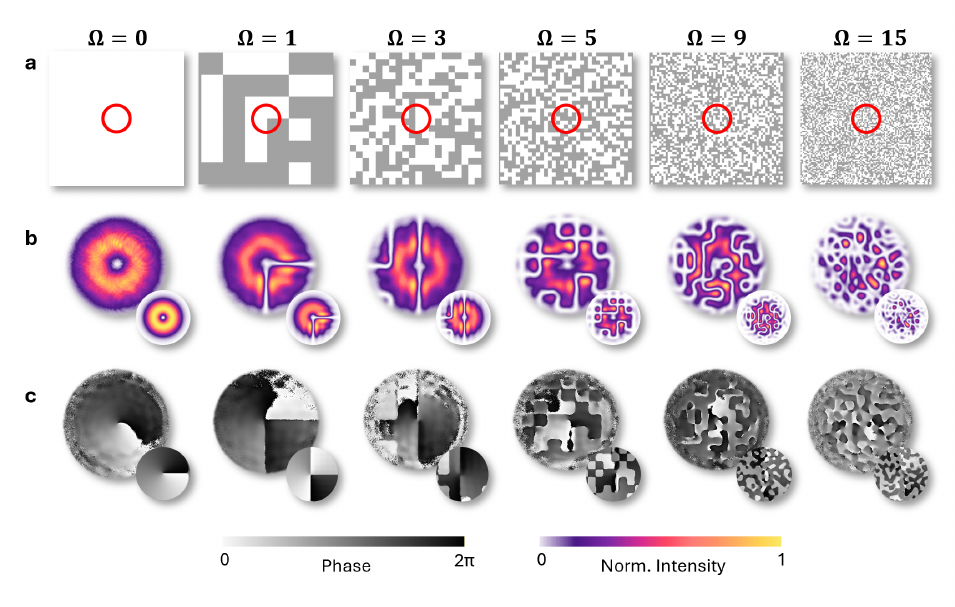}
    \caption{\textbf{Stokes wavefront measurement.} \textbf{a} Binary random phase masks applied to the LG beams on the SLM. Red circles indicate central pixels where the LG beams were superimposed. \textbf{b} Experimentally measured intensity profiles corresponding to $LG_1^0$ for one seed of various strength phase masks in \textbf{a} closely matching simulation insets. \textbf{c} Intramodal phase retrieved from OAM beams closely matching simulation insets.
    \label{fig:suppLGFieldRecons}}
\end{figure*}

\noindent \textbf{Measuring OAM of distorted beams.} Phase measurements are required to measure the OAM of LG beams after the random perturbation is imparted. To retrieve this phase information, we made use of the method proposed in Ref \cite{dudley2014all}, using Stokes polarimetry to perform wavefront sensing. If the vector beam is generated in the horizontal-vertical polarisation basis, the relative phase $\delta$ between the two orthogonally polarised scalar modes can be determined using,
\begin{equation}
    \delta =\frac{1}{2}\arctan{\left (\frac{S_3}{S_2} \right )} \,.
    \label{eq:Intra Modal Phase}
\end{equation}
The hologram generating the vertically polarised component was programmed with the LG beam of desired OAM superimposed with a chosen binary random phase mask. The hologram generating the horizontally polarised component was programmed to generate a Gaussian beam without a binary random phase mask. The horizontally polarised Gaussian beam thus has a flat phase profile, which results in the intramodal phase, $\delta(x,y)$, calculated via Equation \ref{eq:Intra Modal Phase} returning a phase that is identical to that of the vertically polarised beam. This allows us to determine the phase profile of the distorted OAM beam we want to measure. The OAM spectrum $|c_l|^2$ of the perturbed scalar beam $U(x,y)$ was determined numerically from images captured by the CCD using the measured phase profile according to \cite{pinnell2020modal}
\begin{equation}
    |c_l|^2 = \left| \int e^{il\phi}U(x,y) \text{d}x\text{d}y \right|^2,
    \label{Eq: overlap_integral}
\end{equation}
where $U(x,y) = A(x,y)e^{-i\delta(x,y)}$ and $A(x,y)$ is the experimentally determined amplitude measured using the CCD. Figure \ref{fig:suppLGFieldRecons} shows selected experimental results of the measured OAM phase profile using the above-mentioned approach. Figure \ref{fig:suppLGFieldRecons} \textbf{a} shows examples of the binary random phase masks used, \ref{fig:suppLGFieldRecons} \textbf{b} shows the measured intensity of the perturbed LG beams with insets showing the simulated intensity profiles. Figure \ref{fig:suppLGFieldRecons} \textbf{c} shows the retrieved distorted phase profiles with the insets showing the numerically simulated results using the same phase screens. We observe remarkable agreement between the experimental and simulated intensity profiles and close agreement between the experimental and simulated phase profiles, demonstrating the effectiveness of this phase retrieval approach.\\

\begin{figure*}[t!]
    \centering
     \includegraphics[width=0.9\linewidth]{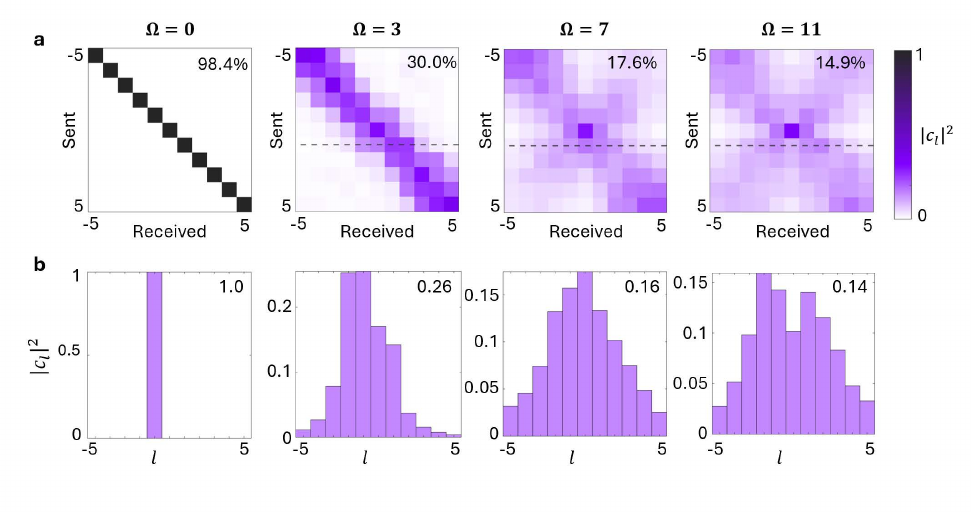}
    \caption{\textbf{OAM crosstalk through random media.}  \textbf{a} Crosstalk matrices measured by modal decomposition into LG modes from $l=-5$ to $l=5$ averaged over 100 random realisations of the digital phase masks for various $\Omega$. Insets quantify the total modal crosstalk.  \textbf{b} Detected average OAM spectrum  of $l=-1$ (black dotted line in \textbf{a}) averaged over 100 random realisations. Insets show the computed fidelities. }
    \label{fig:suppLGdecompPlots}
\end{figure*}

\noindent \textbf{OAM crosstalk through random media.} Using the detected OAM values of the perturbed LG beams we can compute the OAM spectrum. This was used to quantify the severity of the phase masks on the OAM. The OAM crosstalk matrix covering 10 sent and detected OAM values was computed using the overlap integral in Equation \ref{Eq: overlap_integral}. To do this, we performed a modal decomposition into $LG_l^p$ modes with azimuthal indices ranging from $l=-5$ to $l=5$ tested with the same 100 random mask realisations as was applied to the skyrmion numbers. The average crosstalk matrix over these 100 masks is presented in Figure \ref{fig:suppLGdecompPlots} \textbf{a}. Clearly for the  undistorted case ($\Omega =0$) the fidelity is high (98.4\%), however for larger applied distortion strengths, there is more leaking of power into adjacent OAM modes. This spread in OAM is detrimental for information transfer since it becomes unclear at the receiver's end which OAM value was originally sent. We quantify this by computing the modal crosstalk as shown in the inset value in Figure \ref{fig:suppLGdecompPlots} \textbf{a} corresponding to the fraction of the OAM power spectrum along the anti-diagonal. 

Figure \ref{fig:suppLGdecompPlots} \textbf{b} shows the OAM spectrum detected at different distortion strengths for the case where $l=-1$ was sent. This is the average distribution over 100 mask realisations and corresponds to the line-outs of the crosstalk matrices in Figure \ref{fig:suppLGdecompPlots} \textbf{a} indicated by the black dotted lines. These histograms clearly show how the OAM spreads into adjacent modes in agreement with the universal crosstalk law \cite{bachmann2024universal}. The fidelity of the sent $l=-1$ mode similarly decreases as distortion strength increases, as shown by the inset fidelity value. The plots in the main text Figure 5 \textbf{b} correspond to the computed spread of these detected OAM values with the error of detected modes given by the width of the OAM spread fitted with a Gaussian function.\\


\noindent \textbf{Encoding information using topology and OAM.} To illustrate the robustness of the skyrmion wrapping numbers over the OAM values, we encoded an image using a 10 letter alphabet corresponding to 10 OAM values and 10 skyrmion topological wrapping numbers. After passing this through the same 100 realisations of the digital phase masks, we compared the output OAM and skyrmion fidelities. By associating one colour with each skyrmion number or LG mode we used the 10 letter alphabet to encode and send a 10 colour image of a protea flower through digital distortions. This encoding scheme is outlined below. 

First, the original RGB colour image of the protea flower was converted to a 10-colour level image using an inbuilt MATLAB function. These colours were associated with a colour map closely matching the original image as shown in Figure 5 \textbf{c} of the main text. Thus, each pixel in the image only has 1 of 10 possible values corresponding to its colour. This starting colour value represents the information we want to encode in each pixel and transport through the digital random distortion. To perform this encoding, we assign each colour an OAM mode and a skyrmion number. For each distortion strength, we compute the most likely detected mode or skyrmion number using the corresponding 100 crosstalk matrices for the random realisations of the random phase masks shown in Figure 5 of the main text and Figure \ref{fig:suppLGdecompPlots}. The OAM mode with the largest fidelity or the skyrmion number detected and rounded to the nearest integer is the detected output value.  For each pixel in the sent image, the encoded colour value is thus mapped to an output colour value of the most likely detected LG mode or skyrmion number. 
This was repeated for all pixels in the image by traversing the pixels randomly to avoid periodic artifacts arising as a consequence of using 100 mask realisations to encode the original $1024 \times 750$ pixel protea image. To decode this transmitted information and reconstruct the detected image, we map the 10 levels back to the original colourmap forming the images shown in the main text Figure 5 \textbf{d}. The percentages shown in the inset of Figure 5 \textbf{d} are the fraction of correctly encoded colour values in each image. The much greater fraction of correctly encoded topological numbers demonstrated visually by the clearer images thus shows how information can be transferred with high fidelity even under distortion regimes where OAM fails.

\end{document}